\documentclass{article}
\usepackage{amsmath,amsfonts,amssymb}
\usepackage{graphics}
\usepackage{graphicx}
\title{Change detection based on the coefficient of variation in SAR time-series of urban areas }
\author{Elise Colin Koeniguer $^{1}$, and Jean-Marie Nicolas $^{2}$}

\begin{document}
\maketitle


\begin{abstract}
 This paper discusses change detection in SAR time-series. Firstly, several statistical properties of the coefficient of variation highlight its pertinence for change detection. Then several criteria are proposed. The coefficient of variation is suggested to detect any kind of change. 
 Then other criteria based on ratios of coefficients of variations are proposed to detect long events such as construction test sites, or point-event such as vehicles. 
 These detection methods are evaluated first on theoretical statistical simulations to determine the scenarios where they can deliver the best results. Then detection performance is assessed on real data for different types of scenes and sensors (Sentinel-1, UAVSAR). In particular, a quantitative evaluation is performed with a comparison of our solutions with state-of-the-art methods.   
\end{abstract}

\section{Introduction}
\label{sec:intro}

Since the launch of the Copernicus Program, data becomes available on a full, open, and free-of-charge basis. Thus many new services can be developed whether for environmental, civil, industrial, defense, or surveillance applications, using large time-series \cite{geudtner,thepaut}. Urban areas are particularly dynamic: they are being developed at an ever-increasing rate \cite{angel2012atlas}. This article, therefore, focuses on the use of time series of SAR images to be able to image the rate of change during a given observation period in urban areas.

Change detection involves measuring how the attributes of a particular area have changed between two dates \cite{radke}. Among others, change detection makes it possible to monitor natural events, such as flooding events, earthquakes, landslides, but also to observe the urban sprawl. 

The notion of change detection implicitly assumes that we are processing two images: a reference image, considered as the initial case, and a new image, which related to the change to be detected.

Very abundant literature exists on existing algorithms to deal with this problem, from two optical images, or two radar images \cite{bruzzone}. Radar images are very appreciated in this context, by the accessibility to data taking regardless of weather conditions or lighting of the scene, and the statistical properties of signal stability \cite{chatelain}. Unlike optical data, the SAR image is almost insensitive to cloud cover and independent of illumination conditions, and change detection performance is known to be very interesting.

However, it is difficult to be exhaustive on the methods addressing change detection between two SAR images. Let us merely point out that we currently distinguish incoherent detection methods, which exploit only the amplitudes of the images \cite{Vu,gomes}, and coherent methods \cite{washaya2018coherence,preiss2006coherent,liao2008urban}, which also exploit the phase and the concept of interferometric coherence. Furthermore, when the acquisitions are polarimetric, it is necessary to extend the detection criteria to the case of multivariate data \cite{Omati,akbari}.

Easy access to time-series is particularly interesting for urban change detection scenarios. First, because using a higher number of images enables to improve the robustness of single bi-date change detection. Second, because the temporal information can help discriminate different classes of change and thus contribute to the characterization of changes. Indeed, in the context of time-series, the concept of change should be redefined. A signal describing the intensity of a pixel trough time provides a richer source of information and understanding than the notion of “bi-date” change. 

If there is no change inside the scene, the signal is stationary within the entire time series. On the contrary, if the signal is non-stationary, several cases can occur:

\begin{itemize}
\item Durable event: a long process such as building construction. In this case, the temporal signal corresponds to a rising or falling edge. 

\item Occasional events. This can correspond, for example, to the presence of vehicles on a road or a parking lot, or barges on the river. These events are distributed only sporadically.

\item Periodic or quasi-periodic events. It can be the evolution of vegetation which undergoes a seasonal variation.

\item Chaotic evolution. This corresponds for example for crops whose backscattering depends both on meteorological conditions and crop stage.
\end{itemize}
        
Although the use of time series is particularly interesting in this context, less work can be found in the literature on the detection of changes in time series of N dates, because this context of access to data is relatively recent, especially for SAR images. 

Methods of change detection in multivariate time series can be separated into two classes: sequential or online methods, and retrospective or offline analysis. The first ones are  dedicated to the processing of data as they are acquired \cite{mercier2010progressive}, to detect a change point as soon as possible after it occurs. The second ones consider the complete data set at once and look back in time to recognize where a change occurred.
In this paper, we are interested in the offline case: we want to detect in a SAR time series all changes that occurred in a given area and a given period. 

A first strategy consists in considering binary-temporal change detection tests between all possible pairs of dates exhaustively. Resulting tests are represented by a change detection matrix (CDM) \cite{Le2015extraction}
containing all information on changed and unchanged pixels. This matrix is constructed for each spatial position over the time series by implementing similarity cross tests based on the coefficient of variation.
The NORmalized Cut on chAnge criterion MAtrix (NORCAMA) method \cite{su2015norcama} is a spectral clustering methods that consider also the change detection matrix CDM. After a filtering step on the speckle, change criteria between two images are based on a likelihood ratio test using both the original noisy images and the denoised images. Finally the authors apply a change classification by a normalized cut based clustering-and-recognizing method on change criterion matrix (CCM). However, the Change Detection Matrix is a very combinatorial problem, since it requires calculating an NxN matrix in each pixel for a temporal stack of N images. 

Therefore, another alternative is to consider the detection of different ruptures. The omnibus method proposes to test the hypothesis that all polarimetric signals belong to a single statistical population \cite{nielsen2017change}. This global test can be factored into a binary set of tests, which test iteratively that the element $t$ has the same realization polarimetric matrix than the first element of the series. As soon as the test is invalidated, then the test is rejected. It is then possible to consider the following sub-series to search again for another possible rupture. In practice, that means that a subset of binary tests is conducted among all possible pairs corresponding to the first line in the change detection matrix. This approach has been applied on data from different sensors by using a statistical test on polarimetric Wishart distribution in \cite{conradsen2016determining}, \cite{muro2016short}, \cite{rutkowski2018site}.

Finally, the last category of approaches focuses on how to define a criterion to judge the homogeneity of a whole statistical population. It the hypothesis is rejected, then it leads to detect a potential statistical rupture without trying to date it precisely. Among the criteria for detecting activity in time-series, we find in particular:

\begin{itemize}
    \item The MIMOSA method (Method for generalized Means Ordered Series Analysis), detailed in \cite{quin2014mimosa}, is based on the statistical study of the theoretical joint probability density function between the two quadratic and geometric means.  Automatic threshold according to a given false alarm rate, makes it possible to decide whether there is a change or not. 
    
    \item The statistical test involved in the Omnibus method has to decide for a given time profile, whether the variance corresponds to that of a homogeneous statistical population or not. 
    A likelihood ratio test for the homogeneity of several complex polarimetric covariance matrices leads to the ratio of the geometric mean and the arithmetic mean in the single polarimetric case. 
    
  \item The temporal coefficient of variation
(standard deviation x mean $^{-1}$)  is another potentially advantageous candidate  to assess temporal variability because of its simplicity, and its remarkable statistical properties to detect a change. It has been used in \cite{Le2015extraction} and in the visualization method called REACTIV (Rapid and EAsy Change detection on Time-series using the coefficient of Variation). To our knowledge, this is the only visualization method in the literature dedicated to change. In this approach, the critical parameter is the temporal coefficient of variation, which is used to encode the color saturation. Indeed, this parameter has been proven to be particularly sensitive to changes.
  
\end{itemize}


%
In this paper, we therefore propose to build different criteria based on the coefficient of variation for change detection on the entire time-series. The statistical properties of this criterion are at the heart of the strategies for setting the detection functions. They are therefore studied and presented in section 2.

In section 3, we propose, based on these properties, other criteria more specific to distinctive categories of changes; including profiles exhibiting a target appearing only on a single date, typically, a vehicle; and profiles with high returns of longer duration, typically corresponding to the activity of a building site. Section 4 evaluates the three previous detection scenarios by using different simulations.  We present the evaluation protocol, and gives the performances of the methods in section 5, before concluding in section 6

\section{Theoretical considerations for  the coefficient of variation}

\subsection{Generalities}

The coefficient of variation (CV), also known as relative standard deviation, is mathematically defined in probability theory and statistics by the ratio of the standard deviation of the signal by the mean value. Therefore, it is considered as a normalized measurement of the dispersion of a probability distribution. 
In the specific area of SAR images, the coefficient of variation has already been considered for DInSAR (Differential Interferometry SAR) applications.
This technique uses the information from selected points having high backscattering and stable through time, that we call Permanent Scatterer \cite{ferretti}. In this framework, the coefficient of variation is one candidate for the detection of these particularly stable scatterers. 

In \cite{koeniguervisualisation,koeniguervisu}, first theoretical studies have shown that the coefficient of variation is also interesting to detect changes in speckle areas. It has then be used to propose a visualization product that highlights in bright colors all changes that occur in a time-series. 

\begin{figure}
\begin{center}
\includegraphics[width=12cm]{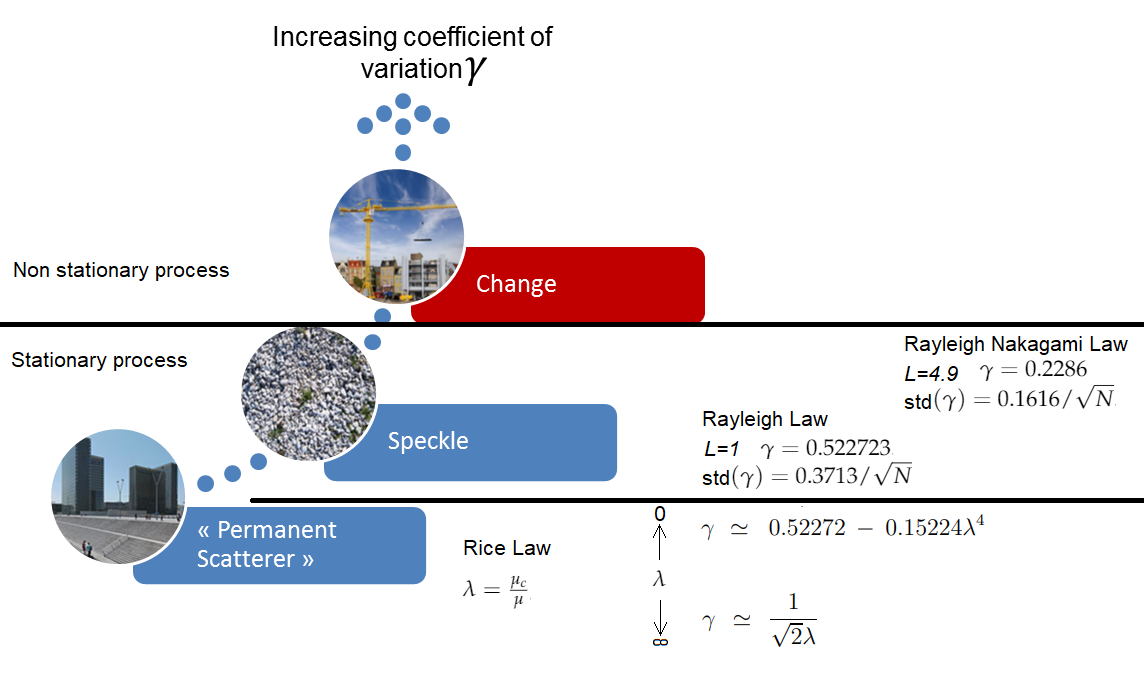}
\end{center}
\caption{Three generic case in temporal profiles with increasing coefficient of variation: the deterministic stable class, the natural non deterministic speckle class, and changes}
\label{fig:cvgeneric}
\end{figure}

It can, therefore, be considered that the coefficient of variation has different statistical properties for at least three categories of temporal profiles: that of a permanent scatterer; that of a natural area of stable speckle, not necessarily correlated in time, but which is stationary; and finally, that of a non-stationary area that we generally interpret as a change. These three general cases are represented in Fig. \ref{fig:cvgeneric}.

The first two of these categories concern cases without any changes. In both these cases, it is possible to derive useful mathematical properties about the coefficient of variations. Also, we discuss these properties in the following paragraphs.

\subsection{The Permanent Scatterer}

In order to identify stationary targets, at least two strategies can be chosen. The first one is to use the interferometric coherence level. The main difficulty relies on the spatial estimation process that implies a spatial resolution loss. The second one relies on the analysis of the amplitude values $A_k$ along the time axis by the so-called dispersion index, which is precisely the coefficient of variation.

The time series of the amplitude values of each pixel containing a permanent scatterer follow the Rice distribution \cite{ferretti, nicolas2019}: 
it corresponds to the superposition of the backscattering of a deterministic target with constant values of phase and amplitude, with an additive fully developed speckle. 
Let $\mu_c$ be the backscattering parameter of the target, then the Rice law can be expressed as: 
If $\mu_c$ is the amplitude of the deterministic signal
and $\mu$ is the parameter of the Rayleigh distribution associated with the speckle noise, the probability associated with a given
value $x$ can be written as: (with $\mu$ and $\mu_c$ $\in \mathbf{R}^+$ ):

$$\mathcal{RC}[\mu_c,\mu](x)=\frac{2x}{\mu^2}e^{(\frac{x^2+{\mu_c}^2}{\mu^2})}I_0(\frac{2\mu_c x}{\mu^2}) $$
where $I_0$ is modified Bessel function of the first kind of order 0. 

Thus, the shape of the Rice distribution depends on the ratio between the amplitude of the deterministic component and the amplitude of the speckle component.

The coefficient of variation $\gamma$ is defined as the ratio of the standard deviation and the average. It can be expressed in terms of the first two statistical moments $m_1$ and $m_2$ of any distribution following:
\begin{equation}
\gamma=\frac{\sqrt{m_2-m_1^2}}{m_1}
\label{eq:cv}
\end{equation}

It is possible to express  $m_1$ and $m_2$ in terms of the law parameters. One of the most elegant ways is to go through the characteristic function of the second-order expressed through Mellin transformations. It makes it possible to obtain the analytic expression of the first moments of the law, by involving the confluent hypergeometric function, also called the Kummer function, denoted by $_1F_1$. 
All mathematical demonstrations are described in \cite{nicolas2018application}. The expressions of the moments thus obtained are not trivial. However, they enable to obtain an analytical expression of the coefficient of variation, either in terms of the Kummer function  $_1F_1$, or in terms of the modified Bessel functions (or hyperbolic Bessel functions) of the first kind $I_0$ and $I_1$: 
\begin{equation}
    \gamma=\sqrt{\frac{4e^{\frac{\mu_c^2}{\mu^2}} {_1F_1}(2;1;\frac{\mu_c^2}{\mu^2})}{\pi(_1F_1(\frac{3}{2};1;\frac{\mu_c^2}{\mu^2}))^2}-1}
\end{equation}
or 
\begin{equation}
    \gamma=\sqrt{\frac{4e^{\frac{\mu_c^2}{\mu^2}}(1+\frac{\mu_c^2}{\mu^2})}{\pi ((1+\frac{\mu_c^2}{\mu^2})I_0(\frac{\mu_c^2}{2\mu^2}) +\frac{\mu_c^2}{\mu^2}I_1(\frac{\mu_c^2}{2\mu^2}))^2}-1}
\end{equation}
These expressions show that the coefficient of variation depends only on the ratio $\lambda=\frac{\mu_c}{\mu}$, i.e., on how the deterministic target emerges from the local noise. 

Moreover, it is possible to find an approximation for the asymptotic cases: 

\begin{itemize}

\item When  $\mu_c$ tends towards infinity, that means a target whose value is infinitely greater than that of the speckle, we find a Normal law  $\mathcal{N}(\mu_c,\frac{\mu}{\sqrt{2}})$, and the coefficient of variation tends towards 0. Moreover, we can find the asymptotic approximation $\gamma=\frac{1}{\sqrt{2}\lambda}$
\item When $\mu_c$ or $\lambda$ tends towards 0, we find the Rayleigh law which followed the fully developed speckle. Then the coefficient of variation will correspond to the case of fully developed speckle $\gamma = 0.52272.$. It is possible to find the asymptotic relation $\gamma = 0.52 - 0.15 \lambda^4 $ for small $\lambda$, satisfactory for $\lambda<0.15$.
\end{itemize}

In order to know how the estimation of the coefficient of variation behaves, we are also interested in the variance of this estimator. Kendall and Stuart in \cite{kendall1977advanced} propose to compute the variance of a function $ g(m_1, m_2) $ by performing a first-order limited expansion of the function around the values $ m_ {0,1} $ and $ m_ {0.2}$. 
This method applied to the coefficient of variation expressed in terms of $m_1$ and $m_2$ in Eq. \ref{eq:cv} leads to find the following formula, as already proposed in \cite {nicolas2006application}:

\begin{equation}
var(\gamma)=\frac{1}{4N} \frac{4m_2^3-m_2^2m_1^2+m_1^2m_4-4m_1m_2m_3}{m_1^4(m_2-m_1^2)},
\end{equation}
where $N$ is the number of images in the sequence, and $m_n$ are the n-th order moment. 

It is of no use to express this variance for the coefficient of variation of Rice's law: the even moments involving exponential functions and the odd moments of the modified Bessel functions of the first kind, no simplification can be expected. However, it has been demonstrated that the moment of order $n$ is proportional to $\mu^n$, and we can deduce from this relation that the variance of the estimator of the coefficient of variation depends only on the parameter $\lambda$ and $N$, and can be written:

\begin{equation}
    var(\gamma)=\frac{1}{N}f(\lambda)
\end{equation}

In all cases, the coefficient of variation associated with a Rice law is statistically lower than that of the corresponding Rayleigh law, and this, especially since the ratio $\mu_c/\mu$ is large.

\subsection{Coefficient of variation for a speckle area without change over time}

A speckle area is typically a forest or area of bare soil. In SAR images, it is commonly accepted that the amplitude of a texture-free speckle distribution follows a Rayleigh-Nakagami law \cite{goodman1976some}:

\begin{equation}
\mathcal{RN}[\mu,L](u)=\frac{2}{\mu}\frac{\sqrt{L}}{\Gamma(L)}(\frac{\sqrt{L}u}{\mu})^{2L-1}e^{-(\frac{\sqrt{L}}{\mu}u)^2}
\end{equation}
where $\mu$ is the shape parameter and $L$ the Equivalent Number of Looks (ENL), and $\Gamma$ is the gamma function.

Let us note that the case of the fully developed speckle corresponds to the case $L = 1 $, and to Rayleigh's law. Rayleigh Nakagami law generalizes the framework of the study to multilooked speckle, as in the case of Sentinel 1 GRD data. Note that 
Rice's law only addresses the fully developed speckle as observed in the SLC data. The case of a deterministic target in multilooked data has not been studied. Its formalism would indeed be even more complicated: it would be necessary to study the spatial averaging of an area containing a deterministic component in a pixel and surrounding pure speckle. This study is out of the scope of this paper. However, it is useful to show, as part of the speckle, how the $L$ parameter can influence the behavior of the coefficient of variation.

Generally,  the parameters of this law are estimated spatially from a homogeneous area. In our approach, we are interested in temporal statistics. We then assume that, for a pixel belonging to a Rayleigh Nakagami law speckle area, and having not perturbed by a change, the various realizations of amplitudes over time of this pixel also follow a Rayleigh Nakagami law. In a way, we consider that the hypothesis of spatial ergodicity can be transformed into a hypothesis of temporal ergodicity. In this case, we will speak in the following of \textbf{stable speckle}.

From Rayleigh Nakagami law, it is possible to derive the expression of empirical moments. We have in particular \cite{nicolas2006application}:

\begin{equation}
m_1=\mu\frac{\Gamma(L+\frac{1}{2})}{\sqrt[]{L}\Gamma(L)}  \textrm{ and } m_2=\mu^2,
\label{eq:mu1}
\end{equation}

which allows to find the following expression of the coefficient of variation, defined as the ratio of the standard deviation and the average:
\begin{equation}
\gamma=\frac{\sqrt{m_2-m_1^2}}{m_1}=\sqrt{\frac{\Gamma(L)\Gamma(L+1)}{\Gamma(L+1/2)^2}-1}
\end{equation}

This expression shows a first interesting property: \textbf{the coefficient of variation will have the same value for all stable speckle zones, whatever the average amplitude of this speckle.} In the particular case $L$ = 1 we find $\gamma=0.522723$.

Knowing the moments of order 1 to 4 of the Nakagami law, it is possible to write the variance of this estimator as a function of $L$:

\begin{small}
{\begin{equation}
var(\gamma)=\frac{1}{4N} \frac{L\Gamma(L)^4 (4L^2\Gamma(L)^2-4L\Gamma(L+\frac{1}{2})^2-\Gamma(L+\frac{1}{2})^2)}{\Gamma(L+\frac{1}{2})^4(L\Gamma(L)^2-\Gamma(L+\frac{1}{2})^2)}
\end{equation}}
\end{small}

In the particular case $L = 1$, we have $var(\gamma) = 0.137881/N$, $\sigma(\gamma) = 0.3713/\sqrt{N}$ respectively for the variance and the standard deviation.

The application of a multilook on a SAR image has the effect of modifying the $L$ parameter. The Sentinel-1 GRD data are delivered with a displayed equivalent number of looks of $L=4.9$, calculated for the theoretical value of the middle of swath and in the middle of orbit \cite{sentinel1}. For a speckle area, this value of $L$ gives $\gamma = 0.2286 $, $\textrm{var}(\gamma) = 0.0216/N$ and $\sigma(\gamma) = 0.1616/ \sqrt{N} $.

This formula shows a second attractive property on this parameter: \textbf{the standard deviation of the coefficient of variation decreases with the number of images $N$ in $\frac{1}{\sqrt{N}} $}.

We can thus imagine that, on temporally stable speckle areas, the coefficient of variation will be constant, with a variance decreasing with the number of images in the sequence.

A last significant result is that the coefficient of variation, calculated on amplitudes according to a Rayleigh Nakagami law, seems to follow a Rayleigh Nakagami law again. This property is still a postulate, verified for many simulations and statistical analyses.

\subsection{Synthesis on the behavior of the Coefficient of Variation}

We have theoretically analyzed the coefficient of variation (CV) for the temporal amplitude profile of a pixel, modeled as the superposition of a fully developed speckle and a deterministic component.

The following properties have been demonstrated:

- The higher the deterministic component, the lower the Coefficient of Variation. It tends to 0 as the relative power between the deterministic component and the speckle increases. It depends only on this ratio.

- When the deterministic component tends towards 0, then CV tends towards 0.52. We are then in the presence of a pure decorrelated speckle. In this case, one finds empirically that the distribution of the CV corresponds to a Rayleigh law.

- In all cases, the variance of the estimator is proportional to $1/N$ where $N$ is the number of images in the time-series.

\section{Change detection criteria}

Thanks to the properties highlighted previously, we propose several decision criteria for the change detection, based empirically on the coefficient of variation.

\subsection{Problem formulation}

Let $S$ be a sequence of $N$ co-registered SAR images acquired at time \{$t_1$ , $t_2$ , . . . , $t_N$ \}. In this paper, a purely temporal approach is proposed. 
That means that we develop criteria that are defined for each pixel independently, by considering only its temporal profile:

$x(t)=\{A(t), t=t_1,...,t_N\}$ where $A(t)$ is the amplitude signal for this pixel in the image acquired at time $t$. 

From this profile, a first intention is to determine if there is a change, of any nature whatsoever, that means with at least one break. 
 A second intention is to detect a specific kind of change. 
However, we do not determine the different break dates, unlike intended in \cite{conradsen2016determining}. 
Therefore, we will propose now the coefficient of variation as \textit{generic change} detection, and other criteria for \textit{specific change} detection. 

Change detection can be formulated as a classical detection problem using a hypothesis testing, with two hypothesis, $H_0=$"no change" against $H_1=$"a change". Then the mathematical expression of the test corresponds simply to the thresholding of the criterion map following:

{\begin{equation}
f(x(t))\overset{H_1}{\underset{H_0}\lessgtr}\lambda
\end{equation}}
with $f$ the criteria function and $\lambda$ the threshold value.

We describe three different functions in the next subsections for generic and specific changes. 
\subsection{Generic change detection}

In section 2, we have illustrated that the coefficient of variation is even lower when the signal is stable. Our first generic change detector is simply this criterion. In this case, function $f$ is defined by: 
{\begin{equation}
    f_1(x(t)) = \gamma(x(t))=\frac{\sqrt{m_2-m_1^2}}{m_1}
\end{equation}}
with 

\begin{equation}
 m_1 = \frac{1}{N}\sum_{t=1}^{N} {x(t)}, ~~m_2 = {\frac{1}{N}\sum_{t=1}^{N}{x^2(t)}}   
\end{equation}

\subsection{Point-event change detection}

We call a point-event or one-time event, a time profile for which a high-backscattering deterministic target is seen at a single date in the profile. In practice, this will correspond to the case of detection of a vehicle. For example, it is the case for boats on the ocean as the probability of having different boats in the same place at different moments is small, even in a long sequence. It is also the case for vehicles in the desert.

The empirical following criterion is proposed. It is defined as the ratio of the coefficients of variation, the first one computed on the profile without the minimal amplitude value, and the other one computed for the profile without its maximal value. The expression is:

{\begin{equation}
f_2(x(t)) = \frac{\gamma({x(t)}_{t\in \{1 \dots N\} \backslash\lbrace{t_{\max}}\rbrace })}{\gamma({x(t)}_{t\in \{1 \dots N\} \backslash\lbrace{t_{\min}}\rbrace})} 
\end{equation}}

where $t_{\min}$ and $t_{\max}$ are the indices on which the signal is minimal or maximal respectively.

Note that a sequential version of this criterion is directly obtained as a special case: it is to consider the ratio of coefficient of variation for the profiles respectively without the first and the last date obtained. This makes it possible to obtain a detector of an event arriving for the last date of acquisition. In this case, the threshold will be applied to: 

{\begin{equation}
f_2'(x(t)) = \frac{\gamma({x(t)}_{t\in \{2 \dots N\}  })}{\gamma({x(t)}_{t\in \{1 \dots N-1\} })}
\end{equation}}

We also propose to compare the $f_3$ criterion with the one computed only on the amplitude means $m$:
{\begin{equation}
f_3(x(t)) = \frac{m_1({x(t)}_{t\in \{1 \dots N\} \backslash\lbrace{t_{\max}}\rbrace })}{m_1({x(t)}_{t\in \{1 \dots N\} \backslash\lbrace{t_{\min}}\rbrace})} 
\end{equation}}

\subsection{Step change detection}

Among the changes that do not belong to the previous category, we will find other categories, such as chaotic changes in radiometry, or the presence of deterministic targets over more extended periods.

The first ones are observed in practice in all agricultural areas. The coefficient of variation is sensitive to these variations related to seasonal changes, moisture changes, or plant growth. These variations can be considered as randomly spread over time. 
Because changes related to agricultural surfaces are not of interest in some applications, we are more interested in the second category of changes. 
Therefore, they correspond to changes related to objects such as infrastructures construction or destruction. The typical profiles generally include steps, that is to say, several dates contiguous with a strong deterministic signal.

We then propose empirical criteria dedicated to this case. The idea is to split the sequence into two parts, to compute the ratio between the coefficient of variation on both sides. 
Then, we take advantage of the temporal redundancy by scanning the different possible cut-off points and by averaging the results thus obtained.

We expect then to have two typical behaviors:
\begin{itemize}
\item If the temporal evolution is simply a radiometric hazard such as those encountered with agricultural surfaces, then the coefficients of variation between the different sub-parts of the profile remain statistically equivalent. It is then expected to obtain a ratio close to 1.
\item On the other hand, in the presence of a change as infrastructure construction or a particular object, several cuts of the profile will lead to finding coefficients of variation significantly different. The ratios should thus have values deviating from 1. 
\end{itemize}

Let $p$ be the variable corresponding to the break date of the sequence. We obtain two sub-parts of the temporal profile noted ${x(t)}_{t\in \{1 \dots p\}}$ and ${x(t)}_{t\in \{p+1 \dots N\}}$. Then the coefficients of variation computed from these profiles are respectively noted $\gamma({x(t)}_{t\in \{1 \dots p\}  })$ and $\gamma({x(t)}_{t\in \{p+1 \dots N\}  })$. To have a significant meaning for the ratio measure, we must respect the statistical constraint for calculating the coefficient of variation, which is to have a sufficient number of images. If we note $M$ the minimum number of images, then $p$ must vary between $M$ and $(N-M)$. 

Then, the function $f$ for the hypothesis test is defined by :

{\begin{equation}
    f_4(x(t)) = 1-\frac{1}{N-2M+1} \sum_{p=M}^{N-M}\min(\frac{\gamma({x(t)}_{t\in \{1 \dots p\}  })}{\gamma({x(t)}_{t\in \{p+1 \dots N\}  })},\frac{\gamma({x(t)}_{t\in \{p+1 \dots N\}  })}{\gamma({x(t)}_{t\in \{1 \dots p\}  })})
\end{equation}}

In this expression, we have a $\min$ operator to have a resulted value between 0 and 1. Moreover, we take one minus the obtained value in order to have low values in case of no change and close to 1 in case of change.  

Once again, we propose to consider the same idea based on mean amplitudes $m_1$ instead of the coefficient of variation:

{\begin{equation}
    f_4(x(t)) = 1-\frac{1}{N-2M+1} \sum_{p=M}^{N-M}\min(\frac{m_1({x(t)}_{t\in \{1 \dots p\}  })}{m_1({x(t)}_{t\in \{p+1 \dots N\}  })},\frac{m_1({x(t)}_{t\in \{p+1 \dots N\}  })}{m_1({x(t)}_{t\in \{1 \dots p\}})})
\end{equation}}

\section{Analysis of performance thanks to statistical simulations}

To evaluate the detection performance based on the previous criteria, we rely on many simulations. These consider different kinds of change. Therefore, we are now analyzing different kinds of profiles.

\subsection{Simulation on one-point events}

\begin{figure}
\begin{center}
\includegraphics[width=8cm]{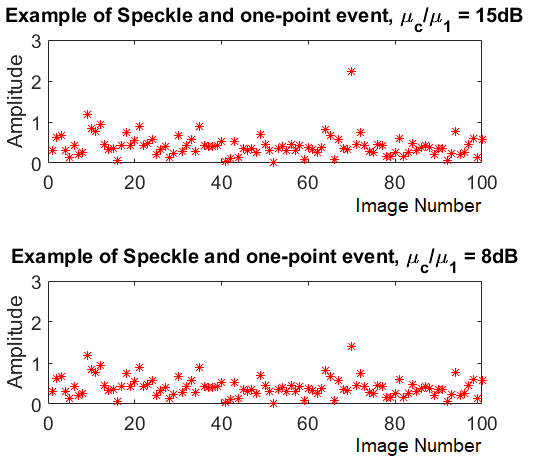}
\end{center}
\caption{Examples of simulated profiles with Rayleigh and one-point event, for two contrast ratios}
\label{fig:profileoneevent}
\end{figure}

We consider here the profiles containing, for the most part, uncorrelated speckle, and for which we have at a given date, the addition of a deterministic target. In this context, we make the number of images of the temporal stack $N$ vary, and the ratio between the amplitude of the target $\mu_c$ and the mean amplitude of the speckle $\mu_1$. Note that $\mu_1$ is not precisely equal to the $\mu$ parameter, but is deduced by Eq. \ref{eq:mu1}. Given the particular dynamics of the radar, we vary this contrast linearly in decibels.
Two examples of such profiles are given in Fig. \ref{fig:profileoneevent}.

In order to estimate the performance of a given criterion, we simulate a large number of profiles.  For each category of changes, we simulate $K=10^6$ reference profiles without any change and $K$ profiles with a change. The investigated criterion is calculated for these two populations. This computation enables to achieve K estimation of this criterion, with and without change. From these K realizations, the empirical distributions of this parameter with and without change are estimated. It is thus classical to compute a Probability of Detection/False Alarm curve for this parameter and this scenario by varying the threshold of the studied criterion.

In the following, we will use colored representations to compare the detection performance of three different criteria obtained by the simulation. Each criterion code a Red, Green and Blue (RGB) sequence of 3 floats in the range 0-1. Zero Probability of Detection for each criteria gives the darkest color, considered the black, and maximum Probability of Detection PD=1 for all criteria gives a white color. When one of the criterion has the strongest probability of detection, the color is a hue near this primary color (red-ish, green-ish, or blue-ish), and when two criteria have the same strongest intensity, then the color is a hue of a secondary color (a shade of cyan, magenta or yellow).

\begin{figure}
\begin{center}
\includegraphics[width=8cm]{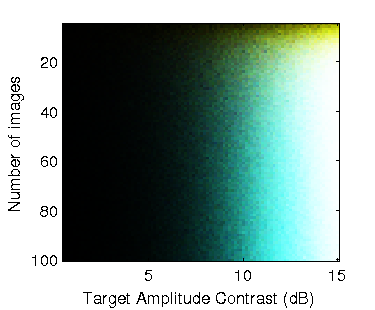}
\includegraphics[width=6cm]{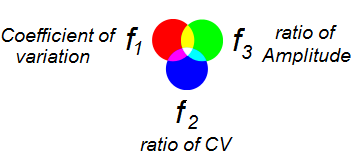}
\end{center}
\caption{Probability of detection of $f_1$,$f_2$,$f_3$, for False Alarm 0.1 \%, for profiles containing a target at a single time, for varying number of images and varying contrasts}
\label{fig:PD1target}
\end{figure}

For the one-point event, we have compared the three criteria $f_1$ (coefficient of variation), $f_2$ (ratio of the coefficient of variation, without min and max value) and $f_3$ (ratio of amplitude means), by fixing PFA to 0.1\%, and 

. In the colored representation of Fig. \ref{fig:PD1target} Red channel corresponds to the probability of detection for the $f_1$ criterion, Green channel for $f_3$ and Blue channel for $f_2$. 

The main findings that we can draw from this figure \ref{fig:PD1target} are as follows:
\begin{itemize}
    \item for $N<20$, the coefficient of variation $f_1$ and the intensity mean ration $f_3$ should be preferred.
    \item For $N>20$, a target can be detected as soon as its contrast is more than 8db. Such a profile example is given in the figure \ref{fig:profileoneevent}.
    \item for $N>20$, ratio criteria $f_2$ and $f_3$ have better performances than $f_1$ for contrasts lying between 8 and 13dB; for contrasts beyond 13dB, all criteria have same performances.
\end{itemize}

\subsection{Mixture of two speckles}

We consider here profiles mixing two different populations of speckle, with a speckle of level given in proportion $P$, and a speckle of higher level, for the rest of the profile (proportion 1-$P$). This type of profile can correspond, for proportions close to one half, to profiles of types of agricultural plots, for which different backscattering levels may appear. Note that when $P$ tends to 0 or 1, we find ourselves in cases of profiles closer those corresponding to one-off events.

In this case, we have again set the number of images to $N$ = 100, we then varied $P$, the percentage of dates at which speckle 1 is present relative to speckle 2 and the ratio between the two average amplitudes of speckle.

\begin{figure}
\begin{center}
\includegraphics[width=8cm]{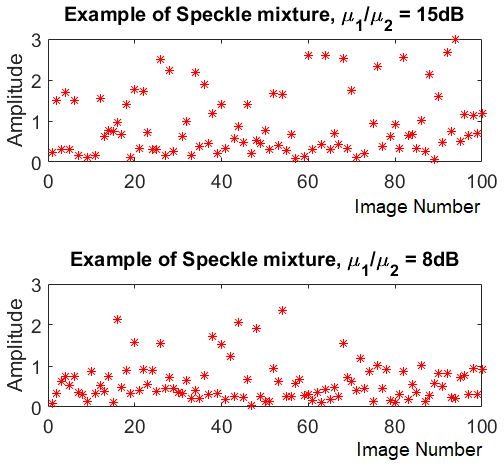}
\end{center}
\caption{Examples of simulated profiles with mixture of two Rayleigh distributions, for two contrast ratios}
\label{fig:profilemixture}
\end{figure}

In the figure \ref{fig:profilemixture}, two examples of profiles are given by setting $P=50$, for two contrast values between speckle: 15dB and 8dB.
These examples illustrate the difficulty of detecting whether a population corresponds in practice to a single Rayleigh population or a mixture of two populations.

\begin{figure}
\begin{center}
\includegraphics[width=8cm]{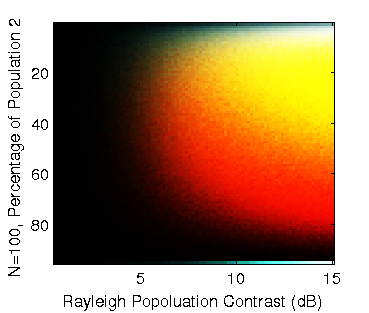}
\includegraphics[width=6cm]{legende1.png}
\end{center}
\caption{Probability of detection of $f_1$,$f_2$,$f_3$, for mixture of two Rayleigh speckle, for False Alarm 0.1 \%, for varying percentage of duration ratios and varying contrasts.}
\label{fig:PD2rayleigh}
\end{figure}
The figure \ref{fig:PD2rayleigh} shows the sensitivity of our different $ f_1 $, $ f_2 $ and $ f_3 $ criteria to this type of mixture profiles. Only the $ f_2 $ criterion is totally insensitive to this mixture, unless it occurs only on one date, which brings us back to the previous case of the point-event target.

The coefficient of variation ($f_1$) remains very sensitive to a mixture of this type. For natural areas with radiometric variations, it will lead to change detection.
The $ f_3 $ criterion remains a little less sensitive than $ f_1 $ but will still detect such events as a change in the case of mixing over low durations.

\subsection{Simulation on P-point events}

We now consider profiles containing speckle, on which a deterministic type target is superimposed for a date proportion equal to $P$. In this case, we have set the number of images to $N = 100$; we make  $P$ vary, the percentage of dates at which the deterministic component is present, and the ratio between the magnitude of the target and the amplitude of the speckle. This type of profile typically corresponds to a site for which there is the presence of a deterministic signal on several dates. It is typical of building construction for example.

\begin{figure}
\begin{center}
\includegraphics[width=8cm]{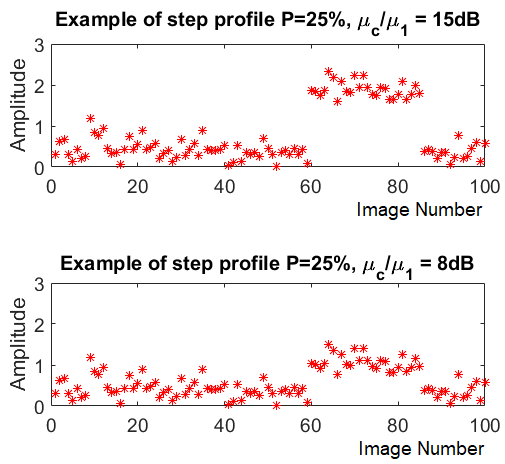}
\end{center}
\caption{Examples of simulated profiles with Rayleigh and one step event, for two contrast ratios}
\label{fig:profilestep}
\end{figure}

The figure \ref{fig:profilestep} illustrates two examples of a profile with a deterministic component over 25\% of the observation duration, for two contrast values.

\begin{figure}
\begin{center}
\includegraphics[width=8cm]{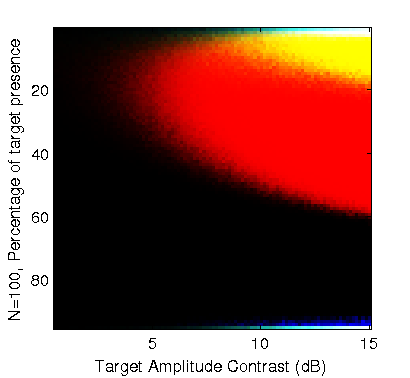}
\includegraphics[width=6cm]{legende1.png}
\end{center}
\caption{Probability of detection of CV and "one-point" detectors, for False Alarm 0.1 \% corresponding to a step signal for varying length of times and contrasts}
\label{fig:PDstep}
\end{figure}

In a first simulation illustrated in figure \ref{fig:PDstep}, probability of detection is represented for the three criteria $f_1$, $f_2$ and $f_3$ for varying proportions $P$ and varying contrasts. These criteria are independent of the order of the values of the profile. 
Results are very similar to the ones presented in figure \ref{fig:PD2rayleigh}.

\begin{itemize}
    \item The criterion $f_2$ is the only one which is specific to a very time-limited change.
    \item The criterion $f_3$ can detect step with duration up to 20\% of total duration. It is of less good specificity. 
    \item The general criterion $f_1$ detect only step with duration less than the half observation period. For longer events, its failure is probably because the deterministic component of the step becomes predominant in the profile, which makes the coefficient of variation decrease again.
\end{itemize}

\begin{figure}
\begin{center}
\includegraphics[width=9cm]{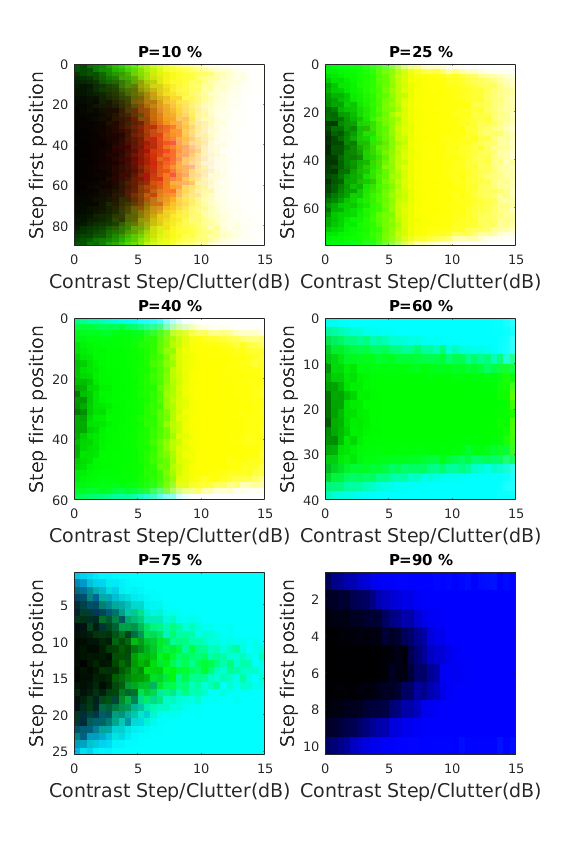}
\includegraphics[width=6cm]{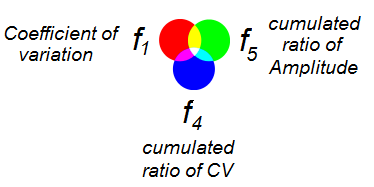}
\end{center}
\caption{Probability of detection of $f_1$, $f_4$ and $f_5$ for False Alarm 0.1 \%, for steps with 6 different lengths of time, for varying step position and contrasts.}
\label{fig:PDstep2}
\end{figure}

In another simulation illustrated in figure \ref{fig:PDstep2}, we consider different fixed values of the proportion $P$ of points that constitutes a step in the profile, and we study the impact of the date of the beginning of this step and the contrast. Probability of detection is represented for the three criteria $f_1$, $f_4$ and $f_5$.

These simulations lead us to the following observations:
\begin{itemize}
    \item Sensitivity to the relative duration of the step ($P$)
    \begin{itemize}
       \item the $f_1$ criterion only manages to detect steps of durations lower than half of the duration of observation.
       \item the criterion $f_4$ can detect only steps with durations greater than half the duration of observation. Beyond a certain duration (P> 80\%), it is the only criterion which manages to detect the step
       \item the criterion $f_5$ is the most versatile compared to the duration of the step.
    \end{itemize}
    \item Sensitivity to the date of the beginning of the step
       \begin{itemize}
           \item only the criterion $f_1$ does not depend, by construction, on the position of the step.
           \item the criteria $f_4$ and $f_5$ depend on this position and detect it all the more difficult since it is centered temporally, and also for particularly brief events ($P$<25\%) or particularly long events ($P$>75\%). 
          \item for high contrast, more than 10 $dB$, there is no more impact on performances
       \end{itemize}
    
\end{itemize}

\section{Performance Evaluation on real SAR data}

Now that our criteria have been evaluated through statistical simulations, we are considering real data validation cases.

\subsection{Choice of test sites and Ground truth}

The first site of interest is chosen on the Saclay area (near Paris, France) between June 2015 and June 2017. Indeed, this site includes many construction sites related to the development of the University Paris-Saclay. Construction dates are available thanks to the local authorities.

\begin{figure}
  \centering
  \includegraphics[width=8cm]{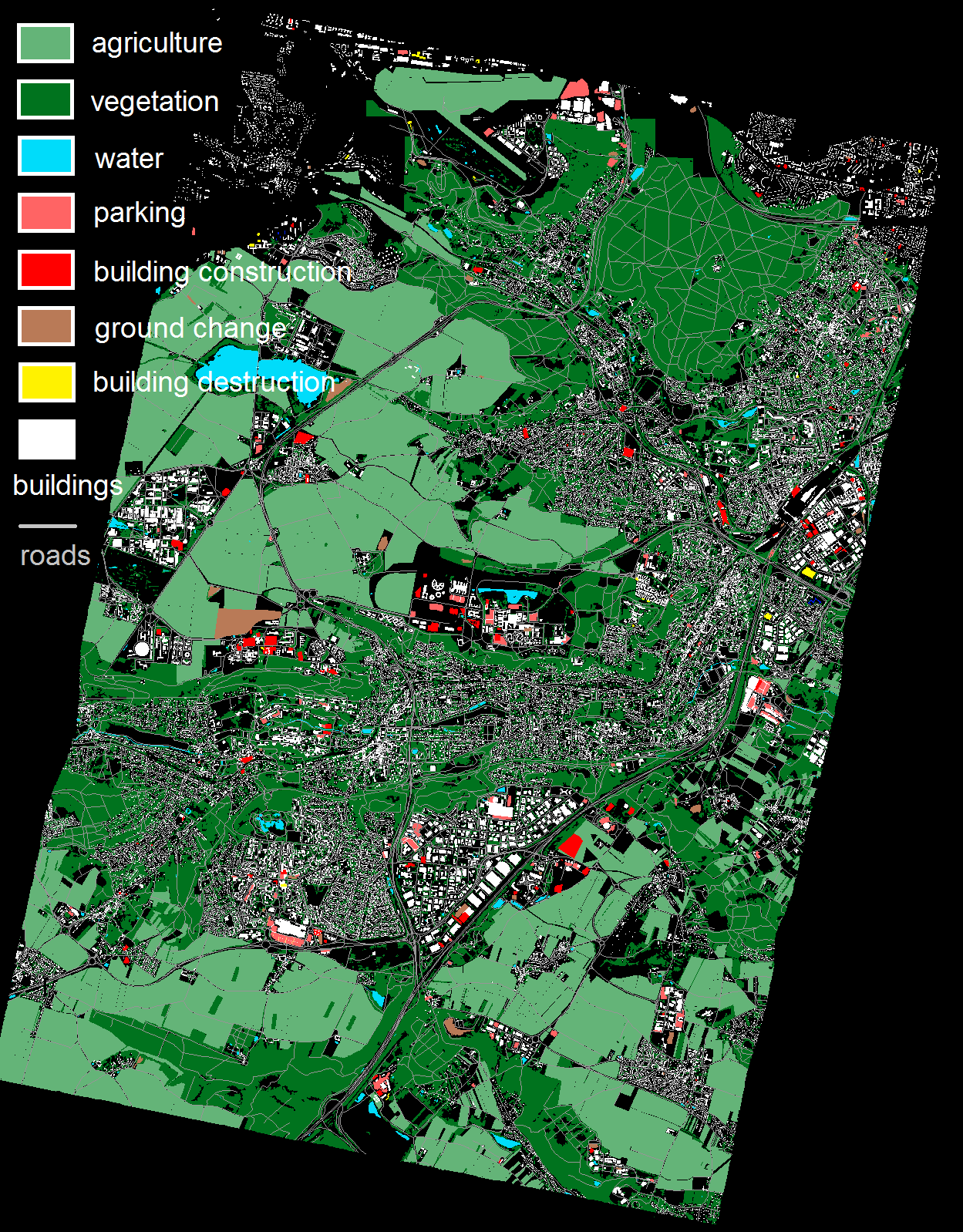}
  \caption{\label{fig_VT} Ground truth established on a test site around Saclay, France}
\end{figure}

The considered area is about 15 km x 12 km. A precise Ground Truth Database has been established, by using two geospatial vector databases:  a topographic database produced by IGN (French National Geographic Agency), and OSM database, corresponding respectively at dates before and after the observed time range.
We kept several standard semantic classes to help interpret the results: water surface, crops, forests, roads,  and buildings.
Both vector files have been converted into raster images by using the GDAL rasterize utility, on a 10m x 10 m common georeferencing grid, and have been subtracted from one another. Then, all the changes found were manually validated or invalidated using past versions of High-Resolution optical images on the Google Earth timeline. The resulting ground truth in Fig. \ref{fig_VT} shows different classes. 
Among them, parking areas, building construction or destruction, and ground changes are assigned to a generic change class used in the performance evaluation. The numbers of change for each class are indicated in table \ref{GT_proportion}.

\begin{table}[h]
\begin{center}
\begin{tabular}{|c|c|c|}
  \hline
  Building 1 & Parking area 2 & Ground area \\
  \hline
  140 & 111 & 42 \\
  \hline
\end{tabular}
\caption{\label{GT_proportion} Number of changes for each category}
\end{center}
\end{table}

This site is observed by the C-band Sentinel-1 sensor.
64 IW Images have been selected during the period in interferometric conditions, in ascending mode. Change detection methods have been applied on L1-SLC (Single Look Complex) images and not on GRD (Ground Range Detected) ones.  We have made this choice for several reasons:

\begin{itemize}
    \item Statistical properties of speckle are preserved in the whole SLC image and do not require to take into account the effect of multilooking.
    
    \item The average phase angles between the polarimetric channels are preserved.
    
    \item in future works, coherent methods could be applied and compared in the present results.
    
\end{itemize}

We have then computed all the criteria maps in the SLC reference grid, before exporting them to the Lambert Conformal Conic projection coordinate system of the Ground Truth. In order to achieve this, the dense non-rigid transformation between SLC and GRD images has been computed thanks to the GeFolki algorithm \cite{brigot2016adaptation}. Then this transformation can be applied to any product computed from the SLC time-series.

In order to take into consideration also one-event targets, a second validation scenario has been considered, with 68 L-band UAVSAR images in the Grizzly Bay, around San Francisco. This dataset is extracted from the SanAnd\_05508 POLSAR stack.

Several boats are available in the images. They belong to the Fifth Reserve Fleet which is docked off the coast of Benicia. The ground truth has been established by manually segmenting the vessels on each image of the stack. We have implicitly assumed that all the salient objects in the image on the water are boats. We have manually checked that they only appear on one date.

For some pixels, an overlap of two boats present at two different dates may occur. In the following, we will estimate the detection performance only on the parts outside these recoveries, to correspond precisely to the "single event" case.

\begin{figure}
  \centering
  \includegraphics[width=6cm]{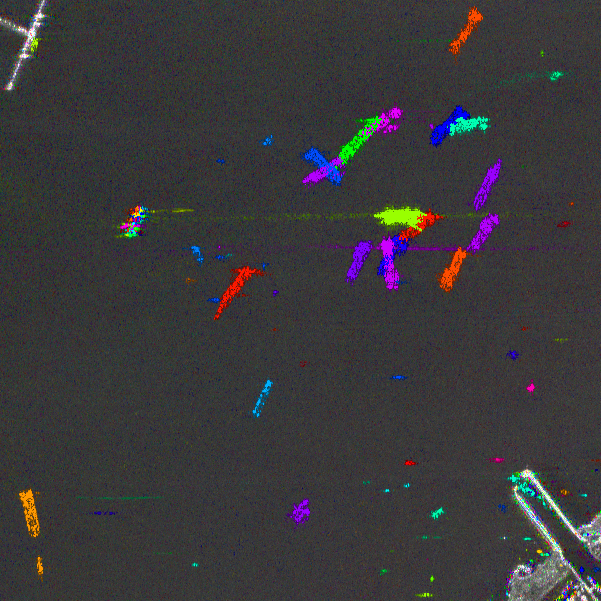}
  \includegraphics[width=6cm]{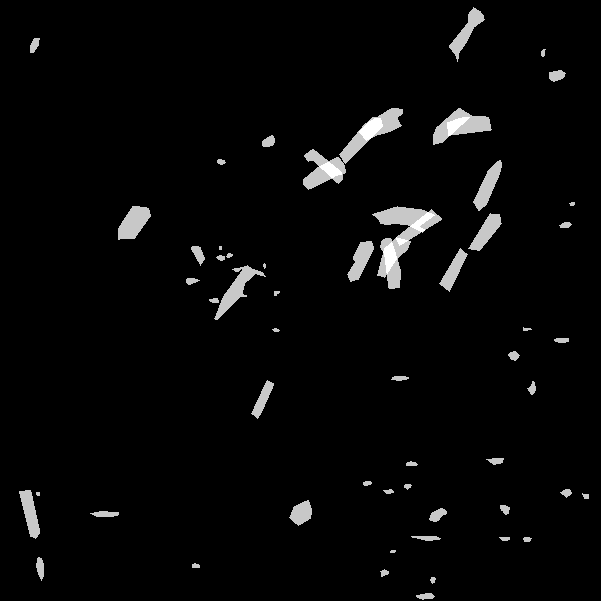}
  \caption{\label{fig_ships} REACTIV method applied to an area of the Grizzly Bay, San Francisco (Left) and corresponding manual data entry for Ground Truth (Right), 600x600 pixels}
\end{figure}

Note that in this data set, we do not strictly have a Ground Truth since we do not use information external to our images; this manual data entry is intended to observe statistical behaviors on real signals.

\subsection{Performance evaluation on Sentinel 1}

The Saclay area contains a significant proportion of cultivated fields and vegetation, in addition to urban areas. The ground truth mainly involves changes in the building elements, some elements of ground modifications and the parking areas.
In this context, we will only evaluate the generic and specific detection of step signals. Besides, we compare our results with those obtained with two other reference methods:

\begin{itemize}
    \item the reference method of the literature on the detection of a change in SAR time series by the so-called "omnibus" method \cite{nielsen2016change},
    \item a state of the art method of change detection by deep learning \cite{daudt2018fully}, applied to Sentinel-2 optical images on the same area. This method is a fully convolutional encoder-decoder paradigm modified into a Siamese architecture, using skip connections to improve the spatial accuracy of the results. It has been trained end-to-end from scratch o Sentinel-2 images, and surpassed the state-of-the-art in change detection, both in accuracy and in inference speed without the use of post-processing. In this paper, it has been tested only on building changes for which it has been trained.
\end{itemize}

The first method consists of a statistical homogeneity test based on a likelihood-ratio test.  Authors propose two versions of this method. A first version considers the test of global homogeneity of the complete temporal profile. It is possible to demonstrate that in this case, the test leads to threshold a criterion proportional to the ratio between geometric mean and arithmetic mean of the intensity signal. When polarimetric information is available, the determinant of the polarimetric covariance matrix replaces the single-channel intensity. 

In a second version of the code, several homogeneity tests are carried out iteratively on subsets of the profile: the intensity considered at date $j$ is compared with the intensity population estimated for all previous dates without breaks. This second version of the code is more comprehensive as it finds the location of all probable breaks. However, it is much more expensive in computing time. In this paper, the first version will be called \textit{omnibus} method and the second version of the \textit{full omnibus} method

Regarding the deep learning method, the comparison must be made with much more caution: the detection of change has been done in a bi-date way, between the start date and the end date. We took care to select images without cloud cover. However, it is likely that many changes over a limited period within the range cannot be seen in this context. On the other hand, the learning has been done mainly on changes relating to new or missing buildings, and therefore, only the performances related to this type of change are considered later. This comparison aims to demonstrate the difficulty of the task of detection of a change from optical images, including with the most advanced algorithms today in deep learning, compared to that of detection from radar images.

The following comparisons have been made by generating Receiver Operating Characteristic (ROC), and Precision-Recall curves. These graphical plots illustrate the ability and precision of our detection parameter as its discrimination threshold is varied.

The ROC curve is created by plotting the Probability of Detection (PD) against the Probability of False Alarm (PFA) at various threshold settings. The Precision-Recall curve 
plots the Precision against the Recall or Probability of Detection (PD) at various threshold settings.

The Probability of Detection (PD) is the percentage of change class objects that are detected. The Recall is equal to Probability of Detection (PD). We consider an object as detected as soon as a minimum of 5\% pixels of the entire object has been detected. We justify this choice by the very sparse nature of a radar signal. For a given building footprint, only a few points are bright points. Thus, we can not hope to detect a break on all the pixels of the considered footprint. 

The Probability of False Alarm (PFA) is given by the percentage of pixels detected that are not lying in the change class of the ground truth. 

Lastly, the Precision is the percentage of pixels declared as detected that corresponds effectively to a change in the Ground Truth. 

\begin{figure}
  \centering
  \includegraphics[width=7.5cm]{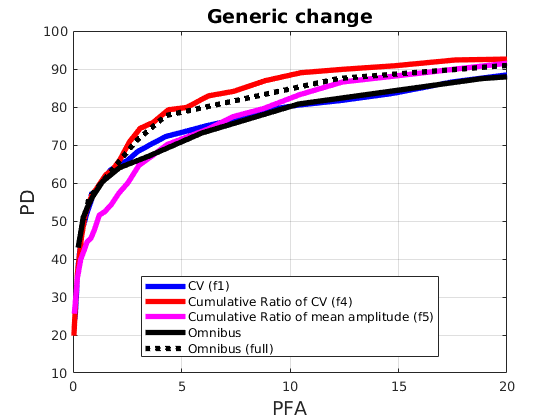}
  \includegraphics[width=7.5cm]{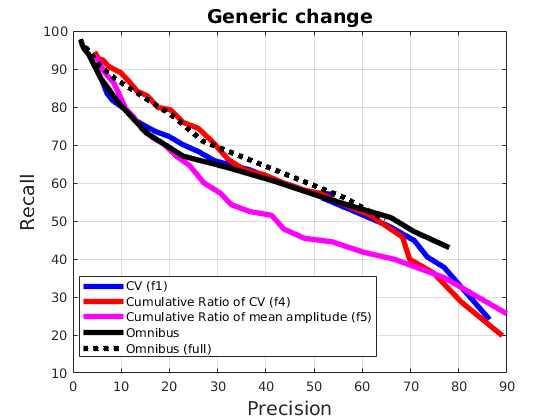}
  \caption{\label{fig_G0_PD} Generic Change Detection (Saclay site, France), PD/PFA on the left, Precision/Recall on the right}
\end{figure}

\begin{figure}
  \centering
  \includegraphics[width=7.5cm]{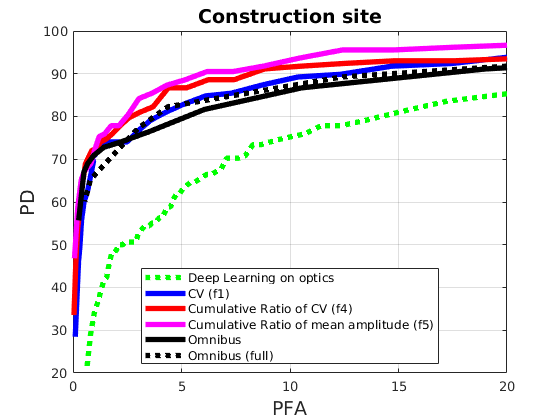}
  \includegraphics[width=7.5cm]{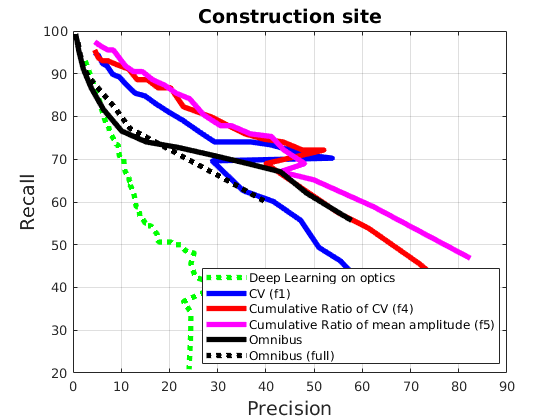}
  \caption{\label{fig_CS0_PD} Case of construction site detection (Saclay site, France) - PD/PFA performances / Precision/Recall performances }
\end{figure}

Figure \ref{fig_G0_PD}  shows the performance for the detection of changes in all classes. Overall, we specify above all that we focus our analysis on low false alarm rates. Indeed, we believe that few operational scenarios can accept 5\% False Alarm, mainly when they are densely distributed among the whole image.
With this point of view, we can not expect a detection rate higher than 70\%. For general changes, the cumulative ratio of CV ($f_4$) and the full omnibus method obtain the best performances. 
Nevertheless, when we restrict the change detection to the building class, the figure \ref{fig_CS0_PD} shows that several methods achieve a much better result, such as 70\% for less than 1\% as False Alarm Rate. This analysis indicates that the main difficulty relies on park areas and ground changes, that cannot be robustly detected on these Sentinel 1 images.

For change detection restricted to buildings in figure \ref{fig_CS0_PD}, except the deep learning method, all methods obtain similar performances for low False Alarm Rates, and two methods are better for higher false alarm rates, that are the cumulative ratio of the coefficient of variation $f_1$ and the cumulative ratio of mean amplitude $f_2$. 

When qualitatively analyzing the criteria maps, we can see that most false alarms are due to high sensitivity to changes in an agricultural area, except for $f_4$. Depending on weather conditions and crop growth, it is likely that the electromagnetic mechanisms and therefore the backscattering levels vary significantly from one date to another. In this case, simulations have proven that cumulative ratio methods $f_4$ are less sensitive to this kind of changes than $f_1$ and $f_3$. On the other hand, the main problem with $f_4$ highlighted by our previous simulations is the difficulty in detecting events shorter than half the duration of observation. Thus, the $f_4$ and $f_5$ criteria have complementary behaviors for comparable performance.

Finally, we note the performance significantly worse with the method of deep learning, highlighting the difficulty to detect changes in optics.

\begin{figure}
  \centering
\includegraphics[width=7.5cm]{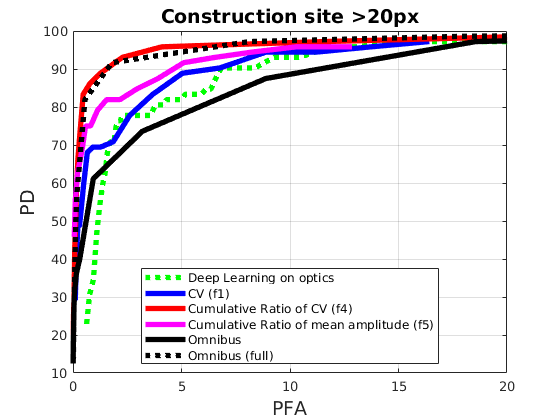}
\includegraphics[width=7.5cm]{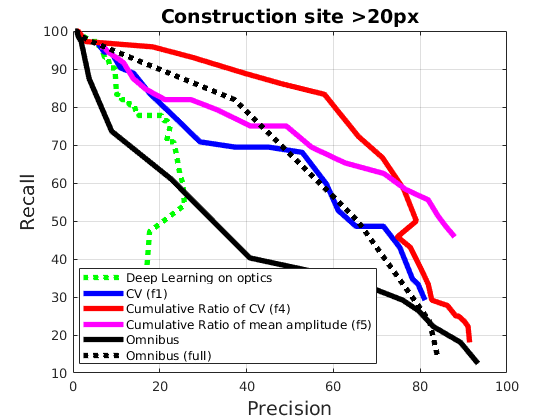}
  \caption{\label{fig_CS20_PD} case of construction site detection of size > 20px (Saclay site, France) PD/PFA performances - Precision/recall }
\end{figure}

In a second step, the performances have been calculated by removing any zone of size less than 20 pixels from the ground truth and the change detection results. This restriction affects mainly the performance on building detection, as shown in figure \ref{fig_CS20_PD} where the improvement is significant, mainly for two methods, the cumulative ratio of CV ($f_4$) and the full omnibus method.

In summary, two methods show excellent performance for change detection on large enough buildings: the cumulative ratio of CV ($f_4$) and the full omnibus method. Nevertheless, remember that the full omnibus method is time-consuming (50 minutes) while the other method is high-speed (4 minutes) for the change detection on radar series composed of 2 (polar) x 64 images of size (1133 x 3205).

\begin{table}[h]
\begin{center}
\begin{tabular}{|c|c|c|}
  \hline
  CV or Omnibus & Ratio of CV or Amplitude & Omnibus (full)\\
  \hline
  6 seconds & 4 minutes & 50 minutes \\
  \hline
\end{tabular}
\caption{\label{TimeCalculation} Time calculation on Intel(R) Core(TM) i7-4710HQ CPU @ 2.50GHz}
\end{center}
\end{table}

\subsection{Performance evaluation on boat detection on UAVSAR}

On this second set of data, we address a very different case from the previous one: that of one-point events, with an example of boat detection.

The performance analysis conducted as previously at the object level would have no meaning here: firstly because some criterion thresholds allow us to obtain a Probability of detection equal to 100\% for a null probability of false alarm; secondly, because we do not have Ground Truth external to the data.

For these reasons, we analyze performances on this real data set, statistically at the pixel scale.

\begin{figure}
  \centering
  \includegraphics[width=8cm]{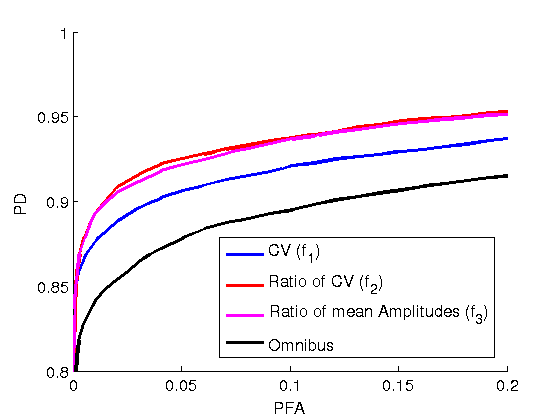}
  \caption{\label{fig_ships_perfs} Performances of change detection for ships at the pixel level, on full polarimetric UAVSAR data set}
\end{figure}

We have selected all the pixels belonging to one single boat without superposition, those represented in gray in the figure \ref{fig_ships}.
The histograms of all these pixels were calculated for the three criteria $ f_1 $ $ f_2 $ and $ f_3 $, and compared to those of a sea area. The ROC curves corresponding to these comparisons of distributions are presented in figure \ref{fig_ships_perfs}.

The $f_2$ and $f_3$ criteria are the ones that lead to the best performances. Be careful, note that the y-axis starts at 80\% detection.
These two specific criteria are statistically better than the generic criteria for the coefficient of variation and the polarimetric omnibus criterion computed on the diagonal covariance matrix.

\section{Conclusion}

In this paper, we have proposed several criteria based on the temporal coefficient of variation to detect change in SAR time-series. 

The coefficient of variation has indeed revealed useful statistical properties: its average values and its variances do not depend on the scale parameter of the speckle. In the case of the presence of a deterministic component, it depends only on the contrast between this component and the clutter. Statistics properties of this criterion have made it a key parameter for deploying robust and extremely fast change detection strategies. 

A first one, the coefficient of variation itself, is dedicated to detection for any change. Also, we proposed to use it as a generic change detector.
Then, since the notion of change corresponds to several types of events, we have proposed other detectors based on ratios, in order to adapt to either short-period event or one-point events such as a vehicle, or more extended events such as construction sites.

These criteria were first analyzed using statistical simulations.
These simulations proved the great genericity of the coefficient of variation to detect any change, but also the specificity of the other criteria to address specific changes. Then these criteria were evaluated on real data.

For step type changes, the specific criteria lead to excellent performance, at least equal to that of the state-of-the-art method "full omnibus" which is much more expensive in computing time. Performances are also much better than those that are currently obtained using a couple of optical data on a similar scenario.
Moreover, the two proposed criteria are complementary: one is likely to respond better to short-steps, in return is more sensitive to changes in all agricultural areas.

For "one-event" changes, the validation has been done on a manual statistical analysis of real high-resolution data containing a large number of ships.

Here again, the two specific criteria lead to the best performances. The simulations tend to say that the criterion based on the CV ratio is even more specific to these short period events. This conclusion allows us to consider using this parameter when detecting a new event with each new acquisition.

In the future, future work will focus on proposing a fusion of the various criteria developed. Ideally, a complete schema could go as far as categorizing change profiles. In parallel, we will analyze the contribution of polarimetric information.

\section*{Aknowledments}
We gratefully acknowledges Behnaz Pirzamanbein for her kindly help on the full omnibus method, and her team for sharing the code. We are very grateful for Rodrigo Daudt, Ph.D. student in Onera, for having tested his algorithm on the Sentinel 2 images over Saclay. We finally thanks for the long-term discussions at the origin of this work our colleagues Beatrice Pinel-Puyssegur and Jean-Michel Lagrange from CEA-DAM. 

UAVSAR data are courtesy NASA/JPL-Caltech, BD TOPO courtesy IGN (France), Sentinel 1 data courtesy ESA. 
The APC was funded by Onera, as part of the MEDUSA Research project.
The formal analysis in this paper has been conducted by E.K. and J-M. N. Conceptualization of this paper, project administration, validation, writing—review and editing have been performed by E.K. 


\section*{Abbreviations}
The following abbreviations are used in this manuscript:\\
\begin{tabular}{@{}ll}
CDM & Change Detection Matrix\\
CV & Coefficient of Variation\\
DInSAR &Differential Interferometry SAR\\
ENL & Equivalent Number of Looks\\
GRD & Ground Radar Detected\\
HSV & Hue Saturastion Value\\
IGN & Institut Géographique National (French National Geographic Agency)\\
MIMOSA & Method for generalized Means Ordered Series Analysis \\
ONERA & Office National D'Etudes et de Recherches Aérospatiales (The French Aerospace Lab)\\
OSM & Open Street Map\\
PD & Probability of Detection\\
PFA & Probability of False Alarm\\
REACTIV & Rapid and EAsy Change detection on Time-series using coefficient of Variation\\
RGB & Red Green Blue\\
ROC & Receiver Operating Characteristic\\
SAR & Synthetic Aperture Radar\\
SLC & Single Look Complex\\
\end{tabular}

\bibliographystyle{plain}
\bibliography{biblio}

\begin{thebibliography}{10}

\bibitem{akbari}
V.~{Akbari}, S.~N. {Anfinsen}, A.~P. {Doulgeris}, T.~{Eltoft}, G.~{Moser}, and
  S.~B. {Serpico}.
\newblock Polarimetric sar change detection with the complex hotelling–lawley
  trace statistic.
\newblock {\em IEEE Transactions on Geoscience and Remote Sensing},
  54(7):3953--3966, 2016.

\bibitem{angel2012atlas}
Shlomo Angel, Alejandro~M Blei, Daniel~L Civco, and Jason Parent.
\newblock {\em Atlas of urban expansion}.
\newblock Lincoln Institute of Land Policy Cambridge, MA, 2012.

\bibitem{brigot2016adaptation}
Guillaume Brigot, Elise Colin-Koeniguer, Aur{\'e}lien Plyer, and Fabrice Janez.
\newblock Adaptation and evaluation of an optical flow method applied to
  coregistration of forest remote sensing images.
\newblock {\em IEEE Journal of Selected Topics in Applied Earth Observations
  and Remote Sensing}, 9(7):2923--2939, 2016.

\bibitem{bruzzone}
L.~{Bruzzone} and D.~F. {Prieto}.
\newblock Automatic analysis of the difference image for unsupervised change
  detection.
\newblock {\em IEEE Transactions on Geoscience and Remote Sensing},
  38(3):1171--1182, 2000.

\bibitem{chatelain}
F.~{Chatelain}, J.~{Tourneret}, and J.~{Inglada}.
\newblock Change detection in multisensor sar images using bivariate gamma
  distributions.
\newblock {\em IEEE Transactions on Image Processing}, 17(3):249--258, 2008.

\bibitem{conradsen2016determining}
Knut Conradsen, Allan~Aasbjerg Nielsen, and Henning Skriver.
\newblock Determining the points of change in time series of polarimetric sar
  data.
\newblock {\em IEEE Transactions on Geoscience and Remote Sensing},
  54(5):3007--3024, 2016.

\bibitem{daudt2018fully}
Rodrigo~Caye Daudt, Bertrand Le~Saux, and Alexandre Boulch.
\newblock Fully convolutional siamese networks for change detection.
\newblock In {\em 2018 25th IEEE International Conference on Image Processing
  (ICIP)}, pages 4063--4067. IEEE, 2018.

\bibitem{ferretti}
A.~{Ferretti}, C.~{Prati}, and F.~{Rocca}.
\newblock {Permanent scatterers in SAR interferometry}.
\newblock {\em IEEE Transactions on Geoscience and Remote Sensing},
  39(1):8--20, 2001.

\bibitem{geudtner}
D.~{Geudtner}, R.~{Torres}, P.~{Snoeij}, M.~{Davidson}, and B.~{Rommen}.
\newblock Sentinel-1 system capabilities and applications.
\newblock In {\em 2014 IEEE Geoscience and Remote Sensing Symposium}, pages
  1457--1460, 2014.

\bibitem{gomes}
N.~R. {Gomes}, P.~{Dammert}, M.~I. {Pettersson}, V.~T. {Vu}, and H.~{Hellsten}.
\newblock Comparison of the rayleigh and k-distributions for application in
  incoherent change detection.
\newblock {\em IEEE Geoscience and Remote Sensing Letters}, 16(5):756--760,
  2019.

\bibitem{goodman1976some}
J.W. Goodman.
\newblock {Some fundamental properties of speckle}.
\newblock {\em JOSA}, 66(11):1145--1150, 1976.

\bibitem{kendall1977advanced}
M.~Kendall and A.~Stuart.
\newblock {The advanced theory of statistics. Vol. 1: Distribution theory}.
\newblock {\em London: Griffin, 1977, 4th ed.}, 1977.

\bibitem{koeniguervisu}
Elise Koeniguer, Jean-Marie Nicolas, and Fabrice Janez.
\newblock {Worldwide multitemporal Ledetection using Sentinel-1 images}.
\newblock {\em Conference on Big Data from Space (BIDS), Munich}, 2019.

\bibitem{koeniguervisualisation}
Elise Koeniguer, Jean-Marie Nicolas, Béatrice Pinel-Puyssegur, Jean-Michel
  Lagrange, and Fabrice Janez.
\newblock {Visualisation des changements sur s{\'e}ries temporelles radar:
  m{\'e}thode REACTIV {\'e}valu{\'e}e {\`a} l’{\'e}chelle mondiale sous
  Google Earth Engine}.
\newblock {\em Revue Française de Photogrammétrie et de Télédétection},
  september 2018.

\bibitem{Le2015extraction}
Thu~Trang Le.
\newblock {\em Extraction d'informations de changement {\`a} partir des
  s{\'e}ries temporelles d'images radar {\`a} synth{\`e}se d'ouverture}.
\newblock PhD thesis, Grenoble Alpes, 2015.

\bibitem{su2015norcama}
Thu~Trang Le, Abdourrahmane Atto, Emmanuel Trouv{\'e}, Akhmad Solikhin, and
  Virginie Pinel.
\newblock {Change detection matrix for multitemporal filtering and change
  analysis of SAR and PolSAR image time series}.
\newblock {\em {ISPRS Journal of Photogrammetry and Remote Sensing}}, page~13,
  june 2015.

\bibitem{liao2008urban}
Mingsheng Liao, Liming Jiang, Hui Lin, Bo~Huang, and Jianya Gong.
\newblock Urban change detection based on coherence and intensity
  characteristics of sar imagery.
\newblock {\em Photogrammetric Engineering \& Remote Sensing}, 74(8):999--1006,
  2008.

\bibitem{mercier2010progressive}
Gr{\'e}goire Mercier.
\newblock Progressive change detection in time series of sar images.
\newblock In {\em Geoscience and Remote Sensing Symposium (IGARSS), 2010 IEEE
  International}, pages 3086--3089. IEEE, 2010.

\bibitem{muro2016short}
Javier Muro, Morton Canty, Knut Conradsen, Christian H{\"u}ttich,
  Allan~Aasbjerg Nielsen, Henning Skriver, Florian Remy, Adrian Strauch, Frank
  Thonfeld, and Gunter Menz.
\newblock Short-term change detection in wetlands using sentinel-1 time series.
\newblock {\em Remote Sensing}, 8(10):795, 2016.

\bibitem{nicolas2006application}
Jean-Marie Nicolas.
\newblock {Application de la transform{\'e}e de Mellin: {\'e}tude des lois
  statistiques de l'imagerie coh{\'e}rente}.
\newblock {\em Rapport de recherche, 2006D010}, 2006.

\bibitem{nicolas2018application}
Jean-Marie Nicolas.
\newblock {Un nouveau formalisme pour la loi de Rice}, 2018.
\newblock
  {https://perso.telecom-paristech.fr/nicolas/jmnicolas\_rice\_2018D003.pdf}.

\bibitem{nicolas2019}
Jean-Marie Nicolas and Florence Tupin.
\newblock A new parameterization for the rician distribution.
\newblock {\em IEEE Geoscience and Remote Sensing Letters}, pages 1--5, 2019.

\bibitem{nielsen2016change}
Allan~A Nielsen, Knut Conradsen, and Henning Skriver.
\newblock Change detection in a time series of polarimetric sar data by an
  omnibus test statistic and its factorization (conference presentation).
\newblock In {\em Image and Signal Processing for Remote Sensing XXII}, volume
  10004, page 1000411. International Society for Optics and Photonics, 2016.

\bibitem{nielsen2017change}
Allan~A Nielsen, Knut Conradsen, Henning Skriver, and Morton~J Canty.
\newblock Change detection in a series of sentinel-1 sar data.
\newblock In {\em Analysis of Multitemporal Remote Sensing Images (MultiTemp),
  2017 9th International Workshop on the}, pages 1--3. IEEE, 2017.

\bibitem{Omati}
M.~{Omati} and M.~R. {Sahebi}.
\newblock Change detection of polarimetric sar images based on the integration
  of improved watershed and mrf segmentation approaches.
\newblock {\em IEEE Journal of Selected Topics in Applied Earth Observations
  and Remote Sensing}, 11(11):4170--4179, 2018.

\bibitem{preiss2006coherent}
Mark Preiss and Nicholas~JS Stacy.
\newblock Coherent change detection: Theoretical description and experimental
  results.
\newblock Technical report, DEFENCE SCIENCE AND TECHNOLOGY ORGANISATION
  EDINBURGH (AUSTRALIA), 2006.

\bibitem{quin2014mimosa}
Guillaume Quin, Beatrice Pinel-Puyssegur, Jean-Marie Nicolas, and Philippe
  Loreaux.
\newblock Mimosa: An automatic change detection method for sar time series.
\newblock {\em IEEE Transactions on Geoscience and Remote Sensing},
  52(9):5349--5363, 2014.

\bibitem{radke}
R.~J. {Radke}, S.~{Andra}, O.~{Al-Kofahi}, and B.~{Roysam}.
\newblock Image change detection algorithms: a systematic survey.
\newblock {\em IEEE Transactions on Image Processing}, 14(3):294--307, 2005.

\bibitem{rutkowski2018site}
Joshua Rutkowski, Morton~J Canty, and Allan~A Nielsen.
\newblock Site monitoring with sentinel-1 dual polarization sar imagery using
  google earth engine.
\newblock {\em Journal of Nuclear Materials Management}, 46(3):48--59, 2018.

\bibitem{sentinel1}
{Sentinel 1 Team}.
\newblock {Sentinel 1 User Handbook}, 2013.
\newblock
  https://sentinel.esa.int/documents/247904/685163/Sentinel-1\_User\_Handbook.

\bibitem{thepaut}
J.~{Thépaut}, D.~{Dee}, R.~{Engelen}, and B.~{Pinty}.
\newblock The copernicus programme and its climate change service.
\newblock In {\em IGARSS 2018 - 2018 IEEE International Geoscience and Remote
  Sensing Symposium}, pages 1591--1593, 2018.

\bibitem{Vu}
V.~T. {Vu}.
\newblock Wavelength-resolution sar incoherent change detection based on image
  stack.
\newblock {\em IEEE Geoscience and Remote Sensing Letters}, 14(7):1012--1016,
  2017.

\bibitem{washaya2018coherence}
Prosper Washaya, Timo Balz, and Bahaa Mohamadi.
\newblock Coherence change-detection with sentinel-1 for natural and
  anthropogenic disaster monitoring in urban areas.
\newblock {\em Remote Sensing}, 10(7):1026, 2018.

\end{thebibliography}

\end{document}